\documentclass[]{emulateapj}

\usepackage{natbib}

\usepackage{longtable}

\def\msun{M$_{\odot}$}

\def\asec{\ifmmode ^{\prime\prime}\else$^{\prime\prime}$\fi}
\def\farcs{\hbox{$.\!\!^{\prime\prime}$}}  	
\def\degs{\ifmmode ^{\circ}\else$^{\circ}$\fi}

\def\GRB{GRB~090926A}
\newcommand{\griz}{$g^\prime r^\prime i^\prime z^\prime$~}
\newcommand{\JHK}{$JHK_S$~}

\shorttitle{The  metal poor DLA of the energetic GRB~090926A}
\shortauthors{Rau et al.}

\begin{document}


\title{A very metal poor Damped Lyman-$\alpha$ system revealed through the most energetic GRB~090926A}


\author{A. Rau\altaffilmark{1},  S. Savaglio\altaffilmark{1}, T. Kr\"uhler\altaffilmark{1,2}, P. Afonso\altaffilmark{1}, J. Greiner\altaffilmark{1}, S. Klose\altaffilmark{3}, P. Schady\altaffilmark{1}, S. McBreen\altaffilmark{1,4}, R. Filgas\altaffilmark{1}, F. Olivares E.\altaffilmark{1}, A. Rossi\altaffilmark{3}, A. Updike\altaffilmark{5}}

\email{arau@mpe.mpg.de}

\altaffiltext{1}{Max-Planck Institute for Extraterrestrial Physics, Giessenbachstrasse 1, Garching, 85748, Germany}
\altaffiltext{2}{Universe Cluster, Technische Universit\"{a}t M\"{u}nchen, Boltzmannstra\ss e 2, 85748, Garching, Germany }
\altaffiltext{3}{Th\"uringer Landessternwarte Tautenburg, Sternwarte 5, D-07778 Tautenburg, Germany}
\altaffiltext{4}{School of Physics, University College Dublin, Dublin 4, Republic of Ireland}
\altaffiltext{5}{Department of Physics and Astronomy, Clemson University, Clemson, SC 29634, USA}

\begin{abstract}

  We   present  VLT/FORS2   spectroscopy  and   GROND  optical/near-IR
  photometry  of   the  afterglow   of  the  bright   {\it  Fermi}/LAT
  GRB~090926A. The spectrum  shows prominent Lyman-$\alpha$ absorption
  with $N_{\rm HI} =  10^{21.73\pm0.07}$\,cm$^{-2}$ and a multitude of
  metal lines  at a common  redshift of $z=2.1062\pm0.0004$,  which we
  associate with the redshift of the GRB. The metallicity derived from
  SII is $\log (Z/Z_\odot)\approx -1.9$, one of the lowest values ever
  found in  a GRB Damped  Lyman-$\alpha$ (DLA) system.   This confirms
  that  the spread  of metallicity  in GRB-DLAs  at $z\approx2$  is at
  least  two  orders of  magnitude.   We  argue  that this  spread  in
  metallicity does  not require a  similar range in abundances  of the
  GRB progenitors, since the neutral interstellar medium probed by the
  DLA is expected  to be at a significant  distance from the explosion
  site.   The hydrogen  column  density derived  from {\it  Swift/XRT}
  afterglow  spectrum  (assuming  $\log (Z/Z_\odot)\approx  -1.9$)  is
  approx.  $\approx100$  times higher  than the $N_{\rm  HI}$ obtained
  from  the Lyman-alpha  absorptions.   This suggests  either a  large
  column  density  of ionized  gas  or  a  higher metallicity  of  the
  circum-burst medium compared to the  gas traced by the DLA.  We also
  discuss  the afterglow  light curve  evolution and  energetics.  The
  absence of a  clear jet-break like steeping until  at least 21\,days
  post-burst   suggests  a   beaming  corrected   energy   release  of
  $E_{\gamma}>3.5\times10^{52}$\,erg,
  indicating that GRB~090926A  may have been one of the most energetic bursts ever detected. \\
 
\end{abstract}

\keywords{gamma rays: bursts -- X-rays: individual (GRB 090926A)}


\section{Introduction}
\label{sec:intro}

For the brief moments of  their existence, Gamma-ray Bursts (GRBs) and
their  X-ray/optical counterparts  are  the brightest  beacons in  the
Universe.   The afterglow  luminosities  and simple  power law  shaped
spectra make  them ideal background lights for  probing the conditions
in their  host galaxies and in intervening  systems through absorption
line spectroscopy.  Although the  afterglows fade away within hours to
days, rapid follow-up observations have been performed for a number of
GRBs  and have  revealed tell tale  features of  the circum-stellar
medium  around  the  progenitor  and  the  interstellar  medium  (ISM)
\citep[e.g.,][]{Castro:2003aa}.   These   studies  provided  otherwise
hidden           details          of           the          kinematics
\citep[e.g.,][]{Klose:2004aa,Fox:2008aa},                     excitation
\citep[e.g.,][]{Vreeswijk:2007aa},     dust     and     gas     content
\citep[e.g.,][]{Savaglio:2004aa,Prochaska:2006aa},   and  the  chemical
abundances    \citep[e.g.,][]{Savaglio:2003aa,Fynbo:2006aa}    in   the
star-forming,      low-mass      galaxies      that     host      GRBs
\citep[e.g.,][]{Le-Floch:2003aa,Christensen:2004aa,Rau:2005ab,Savaglio:2009aa}.

In  this  paper we  present  photometric  and spectroscopic  follow-up
observations of  the bright \GRB,  concentrating on the  energetics of
the burst and the chemical enrichment traced by the optical afterglow.
\GRB\    was    discovered   by    the    Gamma-ray   Burst    Monitor
\citep[GBM;][]{Meegan:2009aa} onboard the  {\it Fermi} Gamma-ray Space
Telescope on  2009 September 26  at $T_0$=04:20:27\,UT and  belongs to
the                         long-duration                        class
\citep[T$_{90}=20\pm2$\,s,][]{Bissaldi:2009aa}.   Further detetections
of   the  prompt   emission   were  reported   from  {\it   Fermi}/LAT
\citep{Uehara:2009aa,Bissaldi:2009ab},        {\it        Suzaku}-WAM,
\citep{Noda:2009aa}  and {\it  Konus}-Wind, \citep{Golenetskii:2009aa}
and  the X-ray  \citep[{\it Swift}/XRT,][]{Vetere:2009aa}  and optical
({\it Swift}/UVOT, Gronwall et al. 2009; Skynet, Haislip et al. 2009a)
afterglows were  also detected. A redshift of  $z=2.1062$ was measured
with VLT/X-Shooter observations \citep{Malesani:2009aa}

Throughout  the  paper, we  adopt  concordance $\Lambda$CDM  cosmology
($\Omega_{\rm  M}=0.27$, $\Omega_{\rm \Lambda}=0.73$,  H$_{0}=71$\ (km
s$^{-1}$) Mpc$^{-1}$), and the convention that the flux density of the
GRB   afterglow    can   be    described   as   $F_\nuν    ∝   \propto
\nu^{-\beta}t^{-\alpha}$.


\section{Observations and Data Analysis}
\label{sec:data}

\subsection{Photometry}
 
Our  follow-up photometry  was initiated  on 2009  September 27.03\,UT
(73\,ks post-trigger) once the  X-ray afterglow had been localized and
its   position   became   accessible   with   the   7-band   Gamma-ray
Optical/Near-Infrared Detector \citep[GROND;][]{Greiner:2008aa} on the
MPI/ESO 2.2\,m  telescope at La Silla,  Chile.  Observations continued
until  28\,days  post-burst (see  Table~4,5  \&  6).   Due to  technical
problems  with  the near-IR  channels,  observations  in  $J$ and  $H$
started only in  the third night, and $K_S$-band  imaging is available
only  from the fourth  night on.   The afterglow  was detected  in all
seven bands  ($g^{\prime} r^{\prime} i^{\prime} z^{\prime}  J H K_S$),
as expected  from the  spectroscopic redshift.  The  best localization
was       measured      to       be      $\alpha_{J2000}$=23:33:36.04,
$\delta_{J2000}$=$-$66:19:26.6   with  a  systematic   uncertainty  of
0\farcs  2 compared to  USNO-B1 \citep{Monet:2003aa},  consistent with
the    Skynet     \citep{Haislip:2009aa}    and    {\it    Swift}/UVOT
\citep{Oates:2009aa} afterglow positions.

\begin{table}
\caption{Log of GROND near-IR photometry \label{tab:grondJHKphot}}
\begin{center}
\begin{tabular}{ccccc}
\hline
\noalign{\smallskip}
$T_{\rm mid} - T_{\rm 0} $ & Exposure [s] & \multicolumn{3}{c}{Brightness$^{a}$}  \\  
$[ks]$ & $[s]$ &   \multicolumn{3}{c}{mag$_{\rm AB}$$^b$}  \\ 
\hline
& & $J$ & $H$ & $K_S$ \\
\hline
265.36 & 120 x 10 &  19.44 $\pm$ 0.04 & 19.24 $\pm$ 0.04 & ... \\
331.98 & 120 x 10 &  19.88 $\pm$ 0.04 & 19.52 $\pm$ 0.04 & 19.27 $\pm$ 0.09 \\
356.28 & 120 x 10 &  20.03 $\pm$ 0.04 & 19.77 $\pm$ 0.07 & 19.55 $\pm$ 0.09 \\
419.06 & 120 x 10 &  20.01 $\pm$ 0.06 & 19.80 $\pm$ 0.07 & 19.57 $\pm$ 0.13 \\
\hline
\end{tabular}
\end{center}
\footnotetext{$^a$Corrected for Galactic foreground reddening and converted to AB magnitudes}
\footnotetext{$^b$For the SED fitting, the aditional error of the absolute calibration of 0.05 ($J$ and $H$) and 0.07 ($K_S$) mag was added quadratically}
\end{table}

The GROND data  were reduced and analyzed with  the standard tools and
methods  described   in  Kr\"uhler  et  al.   (2008).   The  afterglow
photometry was obtained  using point-spread-function (PSF) fitting and
calibrated against  observations of the standard  star field SA115-420
($g^{\prime}r^{\prime}i^{\prime}z^{\prime}$) or against selected 2MASS
stars  \citep{Skrutskie:2006wd}   ($J  H  K_S$).    This  resulted  in
1$\sigma$  accuracies of  0.04\,mag  ($g^\prime z^\prime$),  0.03\,mag
($r^\prime i^\prime$),  0.05\,mag ($JH$), and  0.07\,mag ($K_S$).  All
magnitudes are corrected for Galactic foreground reddening of E$_{\rm
  B-V}=0.03$\,mag \citep{Schlegel:1998ul}.

\subsection{Spectroscopy}

Optical spectroscopy of the  afterglow started on 2009 September 27.08
UT (78\,ks  post-burst) using the  FOcal Reducer and  low dispersion
Spectrograph  2  \citep[FORS2;][]{Appenzeller:1998aa}  mounted on  the
8.2\,m ESO-VLT UT1 telescope in Paranal, Chile (Program ID: 083.D-0903).
We  obtained  $2\times1800$\,s   integrations  using  the  600B  grism
(3600\,\AA -- 6300\,\AA\ coverage) and  a long slit of 1\farcs0 width.
The average  seeing was  $\approx0\farcs75$ resulting in  an effective
spectral  resolution  of  $\approx4.5$\,\AA\  FWHM  at  4500\,\AA,  or
$\approx300$\,km\,s$^{-1}$.

The data were reduced with standard IRAF \citep{Tody:1993aa} routines,
and  spectra  were  extracted  using  an  optimal  (variance-weighted)
method.   Wavelength calibration  was achieved  using HeHgCd  arc lamp
exposures where a fit of  13 arc lines left 0.3\,\AA\ root-mean-square
residuals.   Spectro-photometric  flux  calibration  was  carried  out
against    observations    of   the    standard    star   Hz2    (Oke,
unpublished\footnote{www.eso.org/sci/observing/tools/standards/spectra/hz2.html})
taken   on    2009   September   21.39    UT   \footnote{No   suitable
  spectro-photometric  standard  star  calibration  frames  have  been
  obtained  closer  to  the   time  of  the  afterglow  observation.}.
Corrections for slit losses due  to finite slit width and for Galactic
foreground extinction were applied.


\section{Results}
\label{sec:results}

\subsection{Afterglow Spectral Energy Distribuion \& Light Curve}
\label{sec:grond_results}

\begin{figure}
   \centering
   \includegraphics[angle=-90,width=8.7cm]{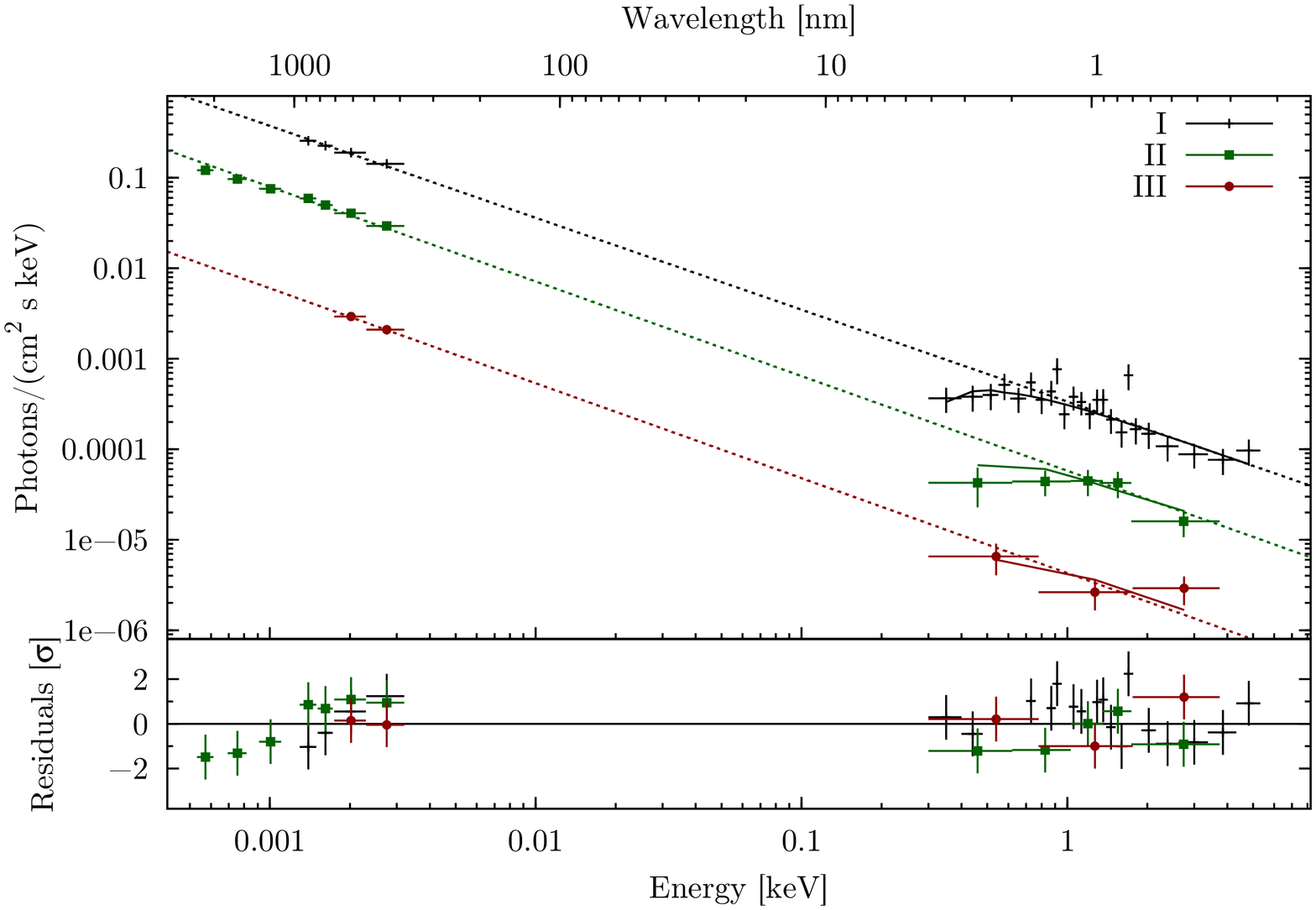}
   \caption{Combined  GROND   and  {\it  Swift}/XRT   spectral  energy
     distribution of the afterglow at $\approx T_0+84$\,ks (interval I
     in  Figure~\ref{fig:grond_lc}), $\approx  T_0+290$\,ks  (II), and
     $\approx  T_0+1.3$\,Ms  (III).  Single  power  laws, modified  by
     N$_H=3.9^{+0.46}_{-0.36}\times10^{21}$\,cm$^{-2}$ at the redshift
     of the burst in addition to the Galactic foreground extinction of
     $2.8\times10^{20}$\,cm$^{-2}$ \citep{Kalberla:2005aa}, fit best.
     The  resulting  slopes   are  $1.02^{+0.03}_{-0.02}$  (I,  90\,\%
     uncertainty,  $\chi^2$=21.3 / 24  d.o.f.), $1.05^{+0.04}_{-0.02}$
     (II, 7.0/10), and $1.04^{+0.08}_{-0.06}$ (III, 2.6/3).  Residuals
     to the fits are shown in the bottom panel.  }
   \label{fig:grond_sed}
   \end{figure}

\begin{figure}
   \centering
   \includegraphics[angle=0,width=8.8cm]{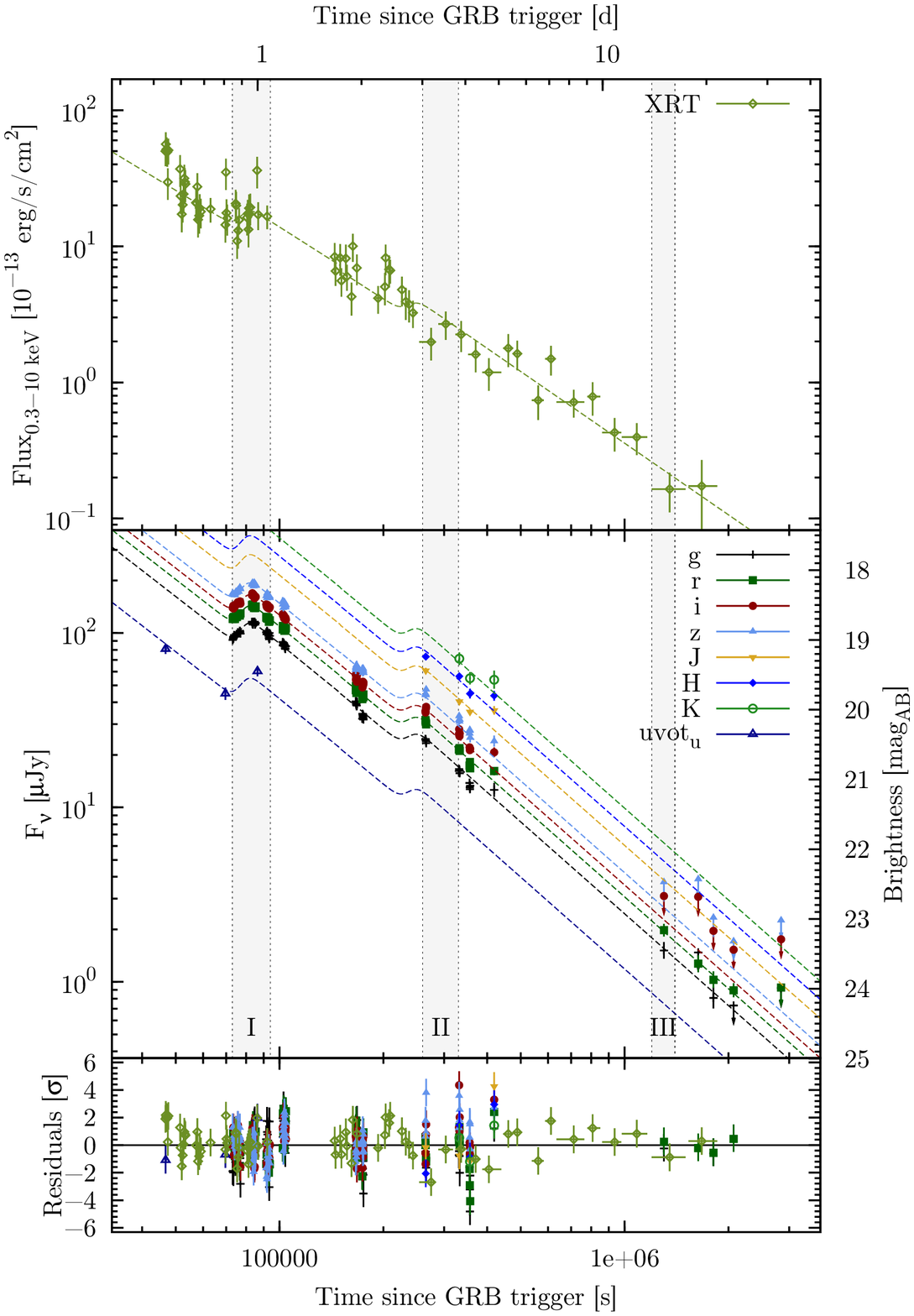}
   \caption{{\it Swift}/XRT (top panel)  and GROND \griz \JHK and {\it
       Swift}/UVOT  $u$-band  (middle)  afterglow light  curves.   The
     dashed lines show a  simplified model consisting of five smoothly
     connected    powerlaws    with    slopes    $\alpha_1=1.6\pm0.2$,
     $\alpha_2=-2.7\pm0.3$,                     $\alpha_3=1.63\pm0.01$,
     $\alpha_5=1.75\pm0.04$          and          break          times
     $T_{1}=T_0+(72.9\pm0.7)$\,ks,        $T_{2}=T_0+(81.7\pm0.3$)\,ks,
     $T_{3}\approx   T_0+180$\,ks,   and  $T_{4}\approx   T_0+250$\,ks
     ($\alpha_4$, the  rising slope leading to the peak around
     $T_{4}$  is  not  constrained   by  our  data).   The  smoothness
     parameters  connecting the power  law segments  were fixed  to 10
     \citep[see  also][]{Beuermann:1999aa}.   Time  intervals for  the
     SEDs   in   Figure~\ref{fig:grond_sed}    are   shown   as   gray
     regions. Residuals are shown in  the bottom pannel.  Data are not
     corrected for Galactic foreground extinction.  }
   \label{fig:grond_lc}
   \end{figure}

Figure~\ref{fig:grond_sed}  and   Figure~\ref{fig:grond_lc}  show  the
optical/near-IR to X-ray spectral energy distributions (SEDs) compiled
at approx. $T_0+84$\,ks, $+290$\,ks, and $+1.3$\,Ms, and the afterglow
light curve, respectively. Here, the Swift/XRT data had been extracted
using the  standard {\it  xrtpipeline} task. The  three SEDs  are best
described  by simple  power  laws with  $\beta_{ox}\approx1.03$ and  a
rest-frame  equivalent   hydrogen  column  densities\footnote{Assuming
  solar                        metallicity.}                        of
N$_H=3.9^{+0.46}_{-0.36}\times10^{21}$\,cm$^{-2}$     assuming     the
abundances of Wilms et al.  (2000).  The spectral slope derived from a
power    law     fit    to    the     optical/near-IR    data    alone
($0.98^{+0.06}_{-0.07}$, $\chi^2$=1.8 for 5 d.o.f.) is consistent with
the  fit including the  X-ray data,  indicating very  little intrinsic
dust reddening.  For the  Small Magellanic Cloud-like dust attenuation
law   \citep{Bouchet:1985aa}  we  find   a  constraint   of  A$_V^{\rm
  host}<0.1$\,mag   at  90\,\%  confidence.    The  absence   of  the
characteristic  2175\,\AA\  bump (would  be  in  the $r^\prime$  band)
similarly restricts a Large Magellanic Cloud or Milky Way-like dust.

For  the light curve  analysis we  simultaneously fit  the 0.3-10\,keV
data  obtained  from  the   {\it  Swift}/XRT  light  curve  repository
\citep{Evans:2007aa}   and  the   GROND   optical/near-IR  photometry,
complemented    by     {\it    Swift}/UVOT    $u$-band    measurements
\citep{Oates:2009aa}.   Here, we  used a  model consisting  of several
smoothly  connected  broken powerlaws  following  the descriptions  of
Beuermann et  al. (1999)  and Liang et  al. (2008).   XRT observations
started  at  $T_0+46.6$\,ks   \citep{Vetere:2009ab}  and  revealed  an
initially decaying  light curve with  $\alpha_1=1.6\pm0.2$, consistent
with the early  UVOT photometry.  This decay slope  is at the boundary
of the distributions of typically found pre- and post-jet-break slopes
\citep{Racusin:2009aa}.  In  the pre-jet scenario  $\alpha_1$ would be
roughly      consistent      with      the      closure      relations
\citep[$\beta=2\alpha/3$,    ][]{Piran:2004aa}    for    the    normal
slow-cooling decay in a constant ISM with the cooling break frequency,
$\nu_c$   \citep{Sari:1998aa},  being   above  the   XRT   band.   The
corresponding  electron index would  be $p=2\beta+1\approx3$.   In the
post-jet-break  case the  closure  relations for  a  constant ISM  and
$\nu_c<\nu$ ($\beta=\alpha/2$) would be similarly well fullfilled.  In
this case, the electron index would lie closer to the canonical value,
$p=2\beta\approx2$.

As reported by Haislip et al. (2009a), around $T_0+$($72.9\pm0.7$)\,ks
the afterglow  brightened again with a  slope of $\alpha_2=-2.7\pm0.3$
before  peaking at  $T_0+(81.7\pm0.3)$\,ks and  then returning  to the
decay with $\alpha_3=1.63\pm0.01$.   A second, similar, re-brightening
episode  likely  occured between  $\approx  T_0+180$\,ks and  $\approx
T_0+250$\,ks.     However,   only    the    subsequent   decay    with
$\alpha_5=1.75\pm0.04$ is covered by  our optical/near-IR data and the
significant  scatter  in  the  X-ray  light curve  prevents  a  better
constraint on this transition period. Deviations from simple power law
light curves are commonly  observed in well-sampled afterglows and can
be  attributed  to a  variety  of  scenarios.   The smooth  rises  and
relatively  constant  (throughout the  observed  period) decay  slopes
suggest that the re-brightnening in  \GRB\ could have been produced by
multiple energy injection episodes,  similar to the model of Bjornsson
et  al.  (2004)  for GRB~021004  \citep[][]{de-Ugarte-Postigo:2005aa}.
Alternative scenarios include processes intrinsic to the forward shock
or   related   to    discontinuities   in   the   circumburst   medium
\citep[e.g.,][]{Lazzati:2002aa}.

Our  data suggest  no  jet-break-like steepening  of  the light  curve
before 21\,days post-burst ($3\sigma$).   However, we can not rule out
a  jet break  prior  to the  start  of the  afterglow observations  at
approx.   $T_0+50$\,ks  ($T_0+16$\,ks   in  the  host  frame).   While
possible, this would be comparably  early. It appears more likely that
XRT   and   GROND  traced   the   normal   spherical   decay  of   the
afterglow. Assuming the simplified model of a uniform jet plowing into
a constant  ISM environment\footnote{Assuming a number  density of the
  ambient  medium  of  1\,cm$^{-3}$  and  a  radiative  efficiency  of
  10\,\%.}   \citep{Sari:1999aa,Burrows:2006aa},  this  implies a  jet
opening  angle   of  $\theta>$9.9\degr\   and  a  beaming   factor  of
$b\approx\theta^2/2>0.015$.    The   10\,keV--10\,GeV   fluence   over
$T_{90}$      was      $(2.47\pm0.03)\times10^{-4}$\,erg     cm$^{-2}$
\citep{Bissaldi:2009ab}, corresponding to  a total isotropic gamma-ray
energy equivalent of E$_{\gamma,iso}= (2.38\pm0.02)\times10^{54}$\,erg
at   10\,keV--10\,MeV  rest  frame   energies.   The   equivalent  jet
angle-corrected         energy         then         amounts         to
$E_{\gamma}>3.5\times10^{52}$\,erg and would  place \GRB\ firmly among
the  most  energetic  bursts known  \citep[see  also][]{Cenko:2010aa},
comparable    to    GRBs    060418    \citep{Cenko:2009ab},    080916C
\citep{Greiner:2009aa}, and 090902B \citep{McBreen:2010aa}.

\subsection{Afterglow Spectrum}

\begin{figure*}[ht]
   \centering
  \includegraphics[angle=0,width=19cm]{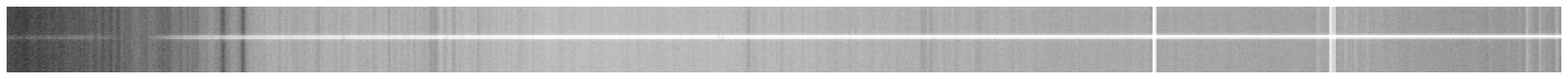}\vspace{-0.2truecm}
  \includegraphics[angle=90,width=19cm]{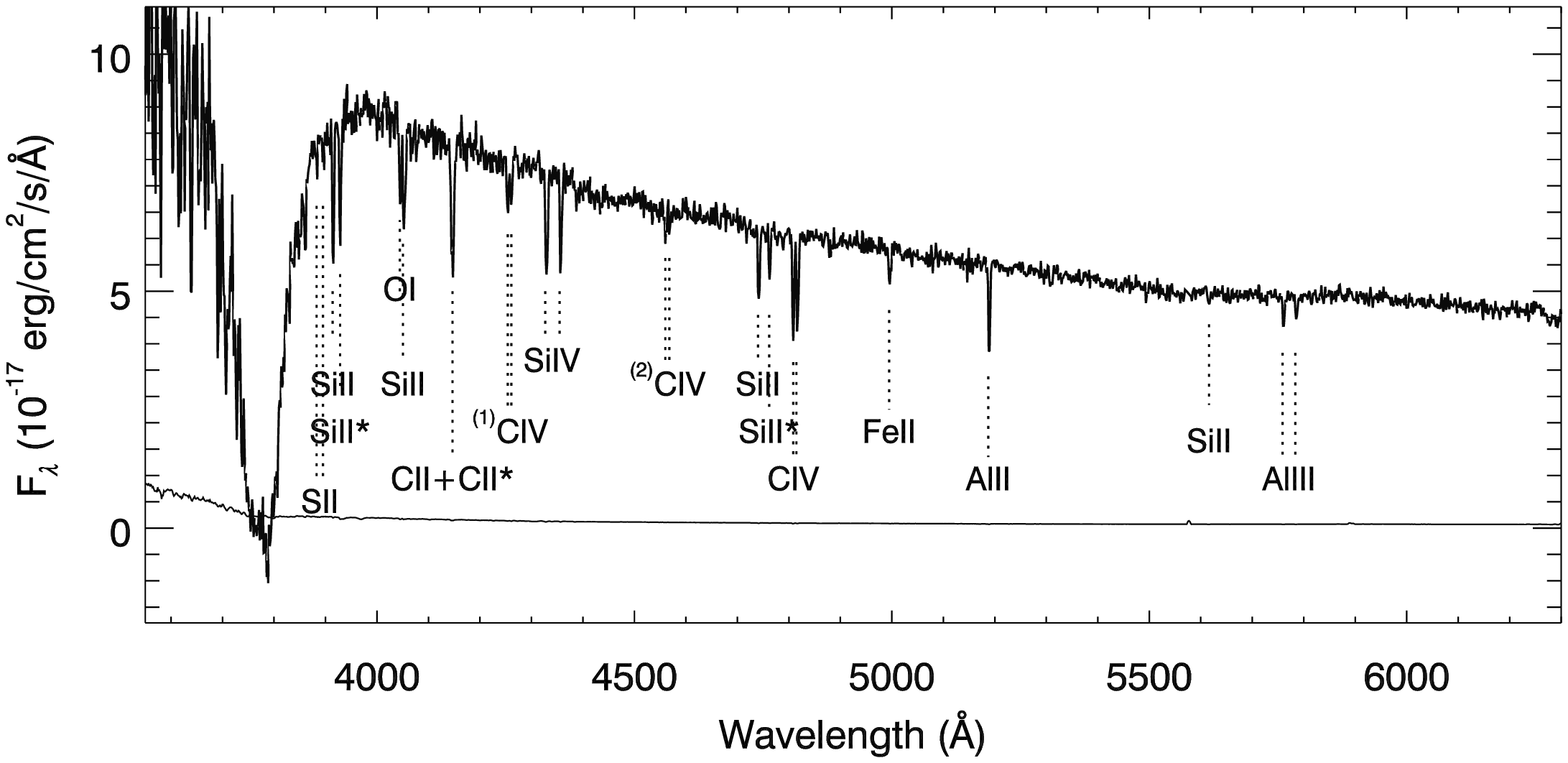}
  \caption{Observer frame VLT/FORS2 spectrum of the afterglow of \GRB\
    obtained 20\,hrs  post-burst. The 2-dimensional  spectrum is shown
    at   the  top   with  telluric   features  remaining,   while  the
    fully-reduced  1-dimensional spectrum is  presented in  the bottom
    panel.   Labels mark  the  most prominent  absorption  lines at  a
    common  redshift of  $z=2.1062\pm0.0004$, as  well as  CIV  in two
    intermediate   systems   at    (1)   $z=1.748\pm0.001$   and   (2)
    $z=1.946\pm0.001$.  The background variance  spectrum is  shown by
    the bottom gray line.}
   \label{fig:spectrum}
   \end{figure*}

\begin{figure}
   \centering
   \includegraphics[angle=0,width=9.1cm]{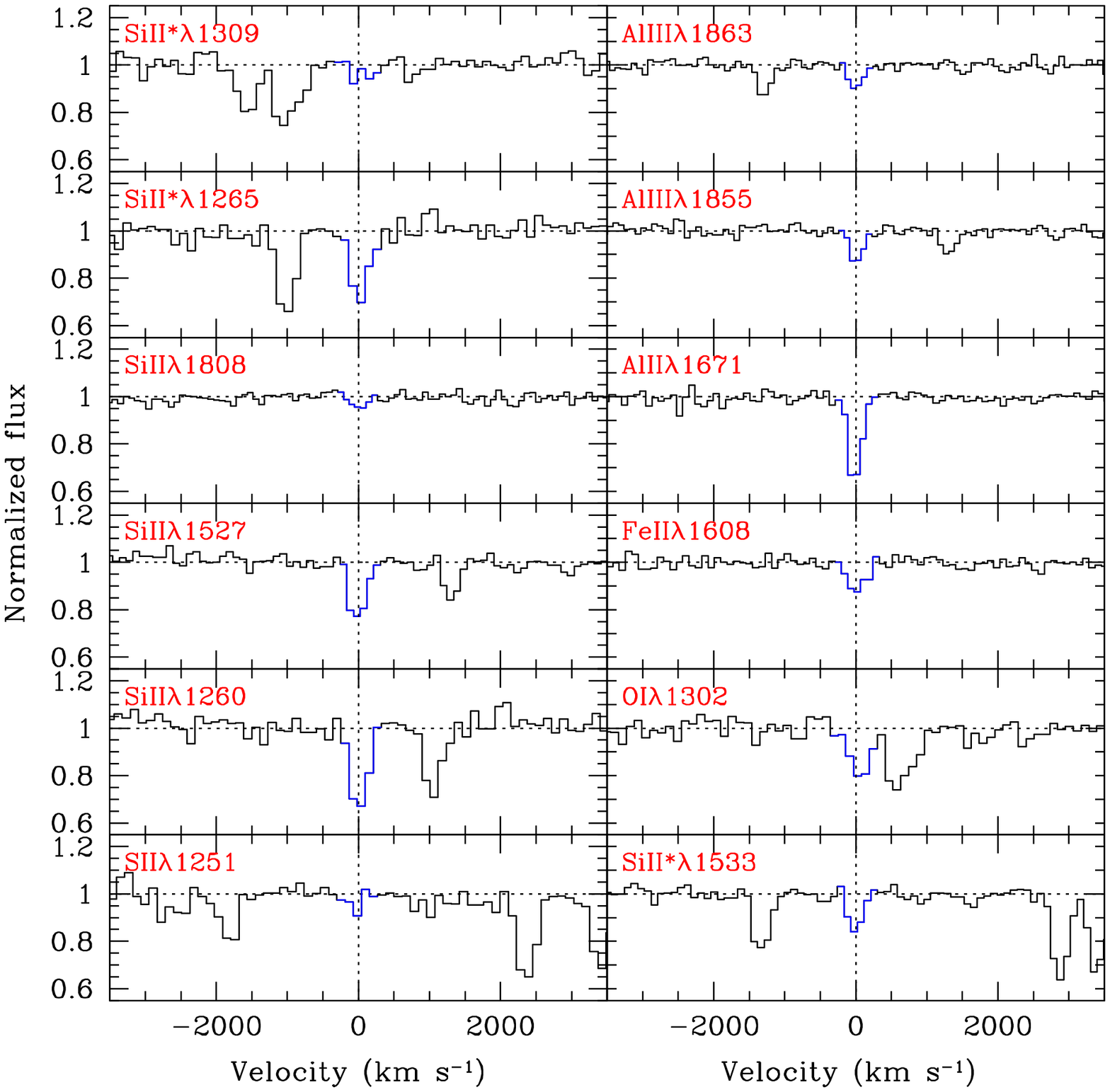}
   \caption{Velocity  profiles  for a  selection  of metal  absorption
     lines. The vertical dotted lines correspond to zero velocity at a
     redshift of $z=2.1062$.}
         \label{fig:lines}
   \end{figure}

The   afterglow   spectrum   (Figure~\ref{fig:spectrum})  displays   a
prominent  Lyman-$\alpha$   absorption  as  well   as  numerous  metal
absorption  lines at  a  common redshift  of $z=2.1062\pm0.0004$  (see
Table~\ref{tab:lines}   and  Figure~\ref{fig:lines}).    Finally,  two
intervening    absorption    systems    at    $z=1.748\pm0.001$    and
$z=1.946\pm0.001$   were   localized  based   on   the  detection   of
CIV$\lambda\lambda1548,1550$               doublets               (see
Table~\ref{tab:intervening}).

\begin{table*}
\caption{Line properties of the  $z=2.1062\pm0.0004$ absorption system.}
\begin{center}
\begin{tabular}{lcccccc}
\hline\hline&&&&&\\[-5pt]
Ion & $\lambda_{\rm obs}$ & $z$ & EW$_r$ & $\log N_X$ & $b$ & [X/H] \\ 
  & (\AA)   & & (\AA)       & [cm$^{-2}$] & (km s$^{-1}$) & \\
[5pt]\hline&&&&&&\\[-5pt] 
HI   & 3775.9   & . . . & . . . & $21.73\pm0.07$ & . . . & . . .  \\
CII$\lambda1334$+CII*$\lambda1335$  & 4146.9& 2.1064  & $0.720\pm0.047$$^a$ & . . . & . . . & . . . \\
AlII$\lambda1670$       & 5189.2 &  2.1059  & $0.461\pm0.035$ & $14.06\pm0.24$ &  . . .$^b$ & $-2.12\pm0.25$ \\
AlIII$\lambda1854$      & 5760.9 &  2.1060  & $0.176\pm0.032$ & $13.25\pm0.08$ & . . .$^b$  & . . . \\ 
AlIII$\lambda1862$      & 5785.9 &  2.1060  & $0.143\pm0.031$ & . . . & . . .  & . . . \\
SiII$\lambda1260$  & 3914.8 &  2.1060  & $0.421\pm0.037$ & $14.97\pm0.26$ & $21.5\pm2.4 $ & $-2.27\pm0.27$ \\
SiII$\lambda1304$       & 4053.0 &  2.1073  & $<0.47$$^c$ & . . . & . . .  & . . . \\
SiII$\lambda1526$       & 4741.8 &  2.1058  & $<0.34$$^c$ & . . . & . . .  & . . . \\
SiII$\lambda1808$       & 5616.8 &  2.1062  & $0.066\pm0.035$ & . . .$^d$ & . . .  &  . . .  \\
SiII*$\lambda1264$ & 3928.5 &  2.1062  & $<0.39$$^c$ & . . . & . . . & . . . \\
SiII*$\lambda1309$      & 4066.0 &  2.1055  & $0.074\pm0.040$ &  $13.71\pm0.13$ & . . .$^b$  & . . . \\
SiII*$\lambda1533$      & 4762.6 &  2.1059  & $0.172\pm0.037$ & . . . & . . .  & . . . \\
OI$\lambda1302$         & 4045.2 &  2.1064  & $0.315\pm0.042$ & $15.40_{-0.27}^{+0.37}$ &    . . .$^b$ &$-3.02_{-0.28}^{+0.38}$ \\
FeII$\lambda1608$       & 4996.2 &  2.1062  & $0.194\pm0.038$ & $14.36\pm0.14$ & . . .$^b$ &  $-2.86\pm0.16$ \\
SII$\lambda1250$   & 3881.9 &  2.1055  & $0.070\pm0.030$ &  $14.96\pm0.15$ &    . . .$^b$ & $-1.89\pm0.16$\\
SII$\lambda 1254$  & 3884.8 &  2.1064  & $0.093\pm0.031$ & . . . &  . . . & . . . \\
NiII$\lambda1317$       & . . . &  . . .   & $<0.114$$^f$  & . . .   & . . .  & . . . \\
NiII$\lambda1370$       & . . . &   . . .  & $<0.219$$^{c,e}$      & . . .   & . . . & . . . \\
NiII$\lambda1741$       & . . . &   . . .  & $<0.110$$^e$ & $<14.1$$^f$  & . . .$^b$ & $<-1.8$ \\
ZnII$\lambda2026$       & . . . &   . . .  & $<0.152$$^e$ & $<13.0$$^f$ & . . .$^b$ & $<-1.2$ \\
[2pt]\hline
CIV$\lambda1548$        & 4808.5 &  2.1059  & $0.511\pm0.040$ & $15.85_{-0.23}^{+0.01}$ &  . . .$^b$ & . . . \\
CIV$\lambda1550$        & 4816.9 &  2.1061  & $0.491\pm0.038$ & . . . & . . .  & . . . \\
SiIV$\lambda1393$       & 4329.5 &  2.1064  & $0.507\pm0.040$ & $14.13_{-0.13}^{+0.20}$ &  $42^{+79}_{-30}$ & . . . \\
SiIV$\lambda1402$       & 4357.0 &  2.1059  & $0.366\pm0.037$ & . . . & . . .  & . . . \\
NV$\lambda 1238$        & . . . &   . . .  & $>0.153$              & $>13.8$ & . . .  & . . . \\
NV$\lambda 1242$        & 3860.9 &  2.1065  & $0.248\pm0.030$       & $14.85_{-0.16}^{+0.21}$ &  . . .$^b$  & . . . \\
[2pt]\hline
\end{tabular}
\end{center}
\footnotetext{$^a$The two lines are blended.}
\footnotetext{$^b$Effective Doppler parameter $b=21.5$  km s$^{-1}$ is fixed.}
\footnotetext{$^c$The line is likely contaminated.}
\footnotetext{$^d$The Apparent Optical Depth method from Savage \& Sembach (1991) gives log$N=15.03\pm0.25$\,cm$^{-2}$ for SiII$\lambda1808$.}
\footnotetext{$^e$3$\sigma$ upper limit.}
\footnotetext{$^f$3$\sigma$ upper limit from EW upper limit.}
\label{tab:lines}
\normalsize
\end{table*}

   \begin{figure}
   \centering
   \includegraphics[angle=-90,width=9.cm]{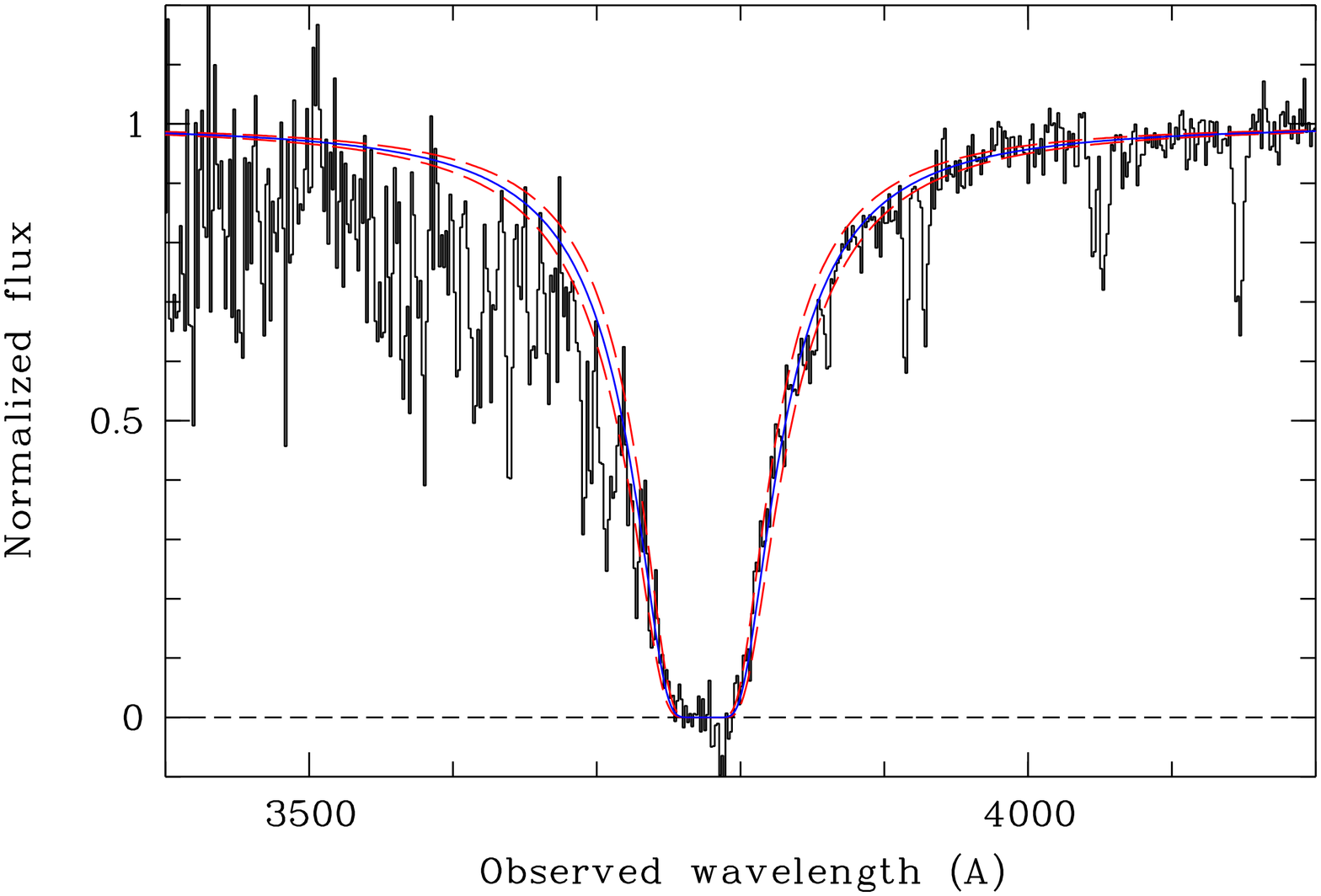}
   \caption{ Lyman-$\alpha$ absorption at the redshift of the GRB host
     galaxy. The  best fit  of the  damped  profile with
     $N_{\rm   HI}  =   10^{21.73\pm0.07}$\,cm$^{-2}$   is  shown   as
     continuous line,  while the dashed contours  indicate the 1-$\sigma$
     uncertainties (see text).}
         \label{fig:lymanA}
   \end{figure}
%

   \begin{figure}
   \centering
   \includegraphics[angle=0,width=8.7cm]{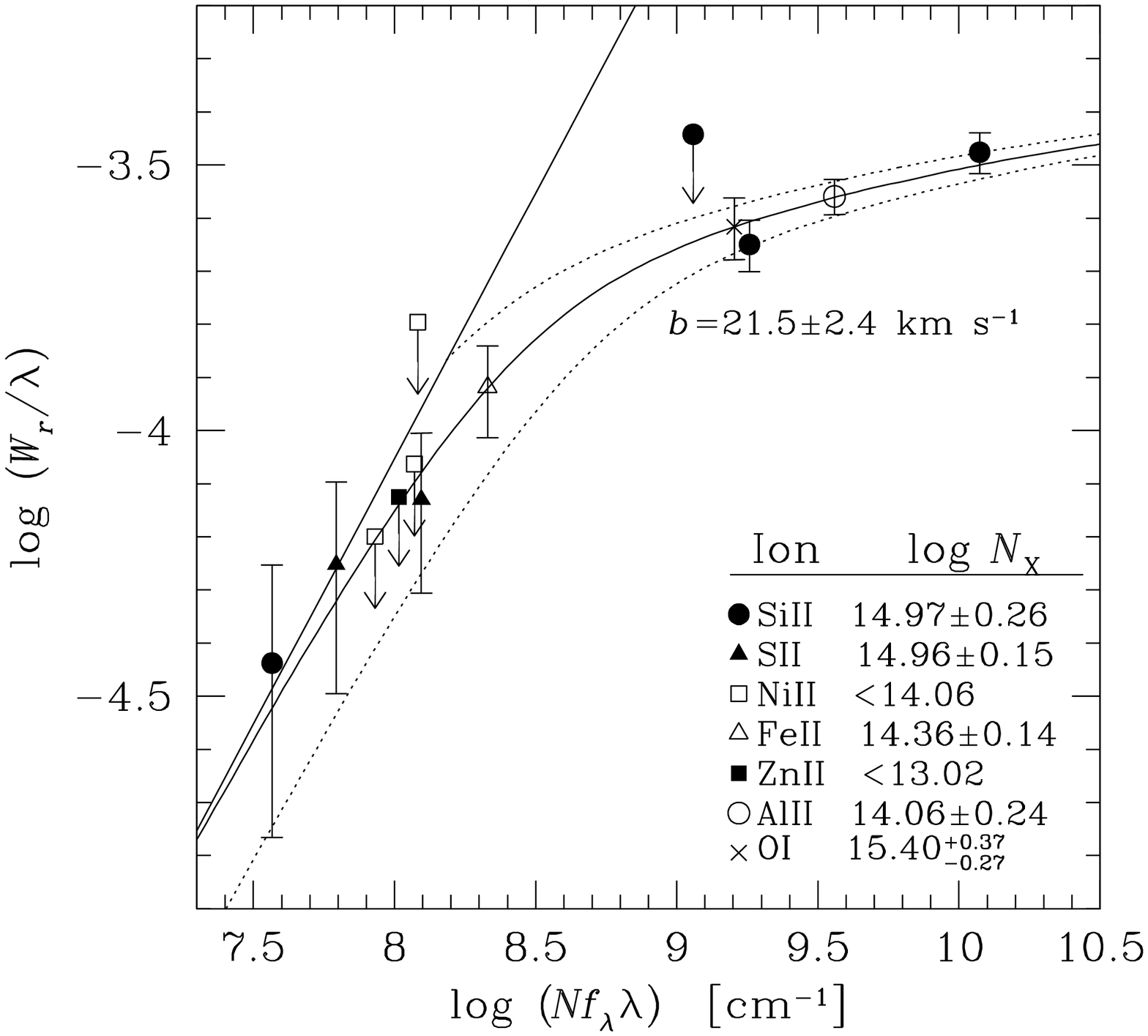}
   \caption{ Curve of  growth of low-ionization species in  the DLA at
     z=2.1062.  The effective Doppler  parameter $b=21.5$  km s$^{-1}$
     has been obtained from the  multiple absorption lines of SiII and
     adopted for all other ions.}
         \label{fig:cog}
   \end{figure}
%

%

\begin{table}
\caption{Intervening absorption systems.}
\begin{center}
\begin{tabular}{lccc}
\hline\hline\\[-5pt]
Ion & $\lambda_{\rm obs}$ & $z$ & EW$_r$ \\ 
       &  (\AA) & & (\AA) \\
[5pt]\hline\\[-5pt] 
CIV$\lambda1548$        & 4254.1 & 1.7477   & $0.192\pm0.039$ \\
CIV$\lambda1550$        & 4261.0 & 1.7478   & $0.175\pm0.039$ \\
\hline
CIV$\lambda1548$        & 4560.7 & 1.9457   & $0.156\pm0.041$ \\
CIV$\lambda1550$        & 4568.1 & 1.9458   & $0.067\pm0.038$ \\
[2pt]\hline
\end{tabular}
\end{center}
\label{tab:intervening}
\normalsize
\end{table}

The column density of neutral hydrogen (HI) was obtained from the best
fit  Voigt  profile  to   the  damping  wings  of  the  Lyman-$\alpha$
absorption feature (solid line in Figure~\ref{fig:lymanA}) letting the
redshift vary in the range  measured from the metal lines.  The dashed
lines  show  the  1-$\sigma$   uncertainties  for  a  fixed  continuum
fit. Systematic effects of the continuum placement may lead to sightly
larger       uncertainties.       With       $N_{\rm       HI}       =
10^{21.73\pm0.07}$\,cm$^{-2}$,  the resulting neutral  hydrogen column
density  is similar  to the  median value  observed in  GRB afterglows
\citep[e.g., ][]{Savaglio:2006aa,Watson:2007aa,Fynbo:2009aa}.

Column densities  of metals were  estimated using the curve  of growth
(COG)  analysis  \citep{Spitzer:1978aa},   and,  if  appropriate,  the
apparent  optical depth (AOD)  method \citep{Savage:1991aa}.   The COG
analysis   is  routinely   applied  when   high-resolution   and  high
signal-to-noise  spectroscopy is not  available. It  provides reliable
results trough the linear approximation when the equivalent width of a
line is EW$ < 0.1$ \AA\  and the effective Doppler parameter is $b>20$
km s$^{-1}$. For stronger lines,  the COG requires the detection of at
least several  lines with different  oscillator strengths of  the same
ion.  The  AOD method is used  for moderate to low  EWs, provided that
the  effective  Doppler parameter  is  not  very low  \citep[$b\gtrsim
10$\,km s$^{-1}$;][]{Prochaska:2007ab}.  As an example, we applied the
linear  approximation  of the  COG  and the  AOD  method  to the  weak
SiII$\lambda1808$ line seen the spectrum of \GRB. The resulting column
density is consistent with the COG analysis of all transitions of SiII
(see Table~\ref{tab:lines}).

For the spectrum shown in Figure~\ref{fig:spectrum}, the best estimate
for    the    effective   Doppler    parameter    comes   from    SiII
($b=21.5\pm2.4$\,km  s$^{-1}$,  Figure~\ref{fig:cog})  which was  also
adopted  for  the  analysis  of  other low  ionization  lines  without
constraints  on  $b$. The  column  densities  derived  using this  $b$
parameter are strictly  correct only for a single  absorbing cloud. If
the  optical-depth  distribution of  absorbers  in  not smooth  (e.g.,
bimodal) the  resulting metal  column densities can  be underestimated
\citep[see     ][for     a     detailed     discussion     of     this
caveat]{Prochaska:2006ab}.   However,  we  note  that  the  $b$  value
derived from  SiII is small  compared to previous measurements  in GRB
afterglow                                                       spectra
\citep[e.g.,][]{Savaglio:2004aa,Berger:2006aa,Vreeswijk:2006aa}.

   In a cloud with $N_{HI} > 10^{20}$\,cm$^{-2}$, the ionization level
   of the  gas is low and  the ionization correction  can be neglected
   \citep{Meiring:2009aa}. In this case, the abundance of an element X
   can be approximated by:

   \begin{equation} {\rm  [X/H]} = \log N_{\rm XII}/N_{\rm  HI} - \log
     (N_{\rm X}/N_{\rm H})_\odot
\end{equation}

where $N_{\rm XII}$  is the column density of  singly ionized elements
with ionization potential just above that of hydrogen\footnote{In some
  cases,  e.g., O  and  N,  already the  ionization  potential of  the
  neutral element, XI, is above  that of hydrogen.  Here, ${\rm [X/H]}
  = \log N_{\rm XI}/N_{\rm HI} - \log (N_{\rm X}/N_{\rm H})_\odot$ can
  be used.}.

The abundances of  Si, Fe, S, Al and O were  found between 0.1\,\% and
1\,\% of the solar  value (see Table~\ref{tab:lines}) with the average
being $\log(Z/Z_\odot)\approx -2.5$.  However, Fe, Al, and Si are
  potentially depleted.  Oxygen suffers from the  saturation region in
  the curve of growth and  its column density is likely underestimated
  \citep[e.g.,][]{Prochaska:2003aa}.    As  Zn   is   undetected,  the
  least-depleted element remaining is  S, which suggests a metallicity
  of $\log(Z/Z_\odot)\approx  -1.9$ for the absorber.  This  is one of
  the  lowest metallicities  measured in  a  GRB Damped-Lyman-$\alpha$
  (DLA) spectrum\footnote{The  current record holder  is GRB~050730 at
    z=3.969         with         $\log(Z/Z_\odot)\approx         -2.0$
    \citep{Chen:2005aa,Starling:2005aa}.}.

Another  interesting finding,  although with  considerable uncertainty
attached to it, is the  apparent overabundance of silicon with respect
to iron ([Si/Fe] $=0.59\pm0.30$), similar to a few previously observed
GRB                                                          afterglows
\citep[e.g.,][]{Prochaska:2007ab,de-Ugarte-Postigo:2009aa}.         One
explanation  for the  high [Si/Fe]  ratio could  be dust  depletion of
iron.   The absence of  significant dust  extinction in  the afterglow
spectral    energy   distribution    makes   this    scenario   rather
unlikely. Alternatively,  this could indicate  an $\alpha$-enhancement
arising from the different  formation time scales of $\alpha$-elements
(short lived  massive stars)  and Fe (type  Ia supernovae).   We note,
that this  is in  potential disagreement with  the low  oxygen content
measured in the spectrum.  However,  as mentioned above, oxygen may be
saturated and its column density thus underestimated.

The spectrum also  shows a number of higher  ionized lines (CIV, SiIV,
and  NV) which  likely arise  from  regions distinct  from those  that
produce  the low-ionization transitions.   Probable locations  for the
hot gas component are within  the star forming regions that hosted the
GRB and/or in the  halo of the host galaxy. In the  case of Si, singly
and  highly-ionized species are  detected in  the spectrum.  Here, the
column  density  of  the  latter  is  low  compared  to  that  of  the
low-ionization transition.


\section{Discussion}
\label{sec:discussion}

\subsection{Burst Energetics}
\label{sec:energetics}

\GRB\ was  likely one of  the most energetic  events in the  sample of
sixteen   bursts   detected   by   both  instruments   onboard   {\it
  Fermi}\footnote{Status  June   20th  2010}.   With   a  jet-angle
corrected energy  release (for  a uniform jet  and a constant  ISM) of
$E_{\gamma}>3.5\times10^{52}$\,erg  (but   see  caveats  described  in
\S~\ref{sec:grond_results})  \GRB\  is  rivaled  only  by  GRB~080916C
\citep[$E_{\gamma}=(3.7\pm0.1)\times10^{52}$\,erg;][]{Greiner:2009aa}
and                           by                           GRB~090902B
\citep[$E_{\gamma}>2.2\times10^{52}$\,erg;][]{McBreen:2010aa}.     This
value  of $E_{\gamma}$  is close  to the  upper bound  allowed  from a
maximally    rotating   neutron   star    in   the    magnetar   model
\citep{Usov:1992aa} and would instead  suggest the remnant of \GRB\ to
be a black hole.

We note  that the inferred jet-angle corrected  energy release depends
not only on whether the afterglow observations traced the pre-break or
post-break  slope,  but  also   on  the  assumption  for  the  density
distribution of  the circumburst medium  and for the jet  geometry.  A
wind-shaped  environment\footnote{With  a   progenitor  mass  loss  of
  $10^{-5}$\,M$_{\sun}$  yr$^{-1}$  and   wind  velocity  of  1000\,km
  s$^{-1}$ \citep{Chiosi:1986aa}.}  would  decrease the lower limit on
the  jet angle  and  thus  relax the  requirement  on $E_{\gamma}$  to
$>4.1\times10^{51}$\,erg;      \citep[][]{Livio:2000aa,Greiner:2003aa}.
However, the spectral and temporal  slopes are in disgreement with the
closure     relations     predicted     for     the     wind     model
\citep{Chevalier:2000aa}     and     favour     a     constant     ISM
\citep{Piran:2004aa}.  Alternatively,  $E_{\gamma}$ can be  reduced by
adopting  a more  realistic geometry  than the  top-hat  (uniform) jet
model  \citep[e.g.,][]{Rhoads:1997aa,Panaitescu:1999aa}.  A structured
jet     \citep{Wijers:1999aa}     or     a     multi-component     jet
\citep[e.g.,][]{Berger:2003aa}   would    for   instance   allow   the
$\gamma$-ray   emission   to    be   tighter   collimated   than   the
long-wavelength counterpart.  A  similar discussion has been presented
by  Cenko et al. (2009)  for five  energetic  pre-{\it Fermi}  bursts,
indicating that in fact our  understanding of the jet geometry in GRBs
and their energetics  is still incomplete. More insights  into the jet
structure    require   an    increasing   number    of   well-sampled,
multi-wavelength afterglow light curves, such as routinely produced by
{\it Swift} and GROND.   Radio observations with the upcoming Expanded
Very  Large  Array  are   expected  to  provide  the  first  realistic
constraint on $E_{\gamma}$ for a larger sample of bursts.

\subsection{Neutral Hydrogen Column Density}

The $N_{\rm HI}$ derived from the Lyman-$\alpha$ absorption is similar
to the equivalent $N_{\rm H}$ estimated from the soft X-ray absorption
(log$N_{\rm H}\sim21.6$;  see \S~\ref{sec:grond_results}).  The latter
is dominated  by $\alpha$-chain elements  and thus mainly a  proxy for
the oxygen  column density assuming solar metallicity.   If the oxygen
abundance of the gas that  causes the X-ray absorption were similar to
that  in  the  DLA  (see Table~\ref{tab:lines},  then  the  equivalent
$N_{\rm  H}$ would  be approximately  two orders  of  magnitude larger
(log$N_{\rm  H}\sim23.7$).  While  there is  generally  no correlation
between  X-ray  and  optical   column  densities  for  GRB  afterglows
\citep{Watson:2007aa},   the  $\approx100$  times   larger  equivalent
$N_{\rm H}$  from the X-ray absorption compared  to the Lyman-$\alpha$
profile fit  for \GRB\  would be  one of the  most extreme  cases.  It
would require  the large amount  of hydrogen missed  by the DLA  to be
ionized.   The  column  density  of  log$N_{\rm  H}\approx23.7$  would
correspond  to   a  surface  density   of  $\approx4\times10^3$\,\msun
pc$^{-2}$, comparable to the  H$_2$ surface densities of circumnuclear
starbursts   \citep{Kennicutt:1998aa}.   Alternatively,   the  surface
density can be  reduced if the metallicity of the  gas in the vicinity
of the burst is higher that that traced by the DLA further out.

\subsection{Metallicities in GRB Damped-Lyman-$\alpha$ Systems}

\begin{figure}
   \centering
   \includegraphics[angle=0,width=8.7cm]{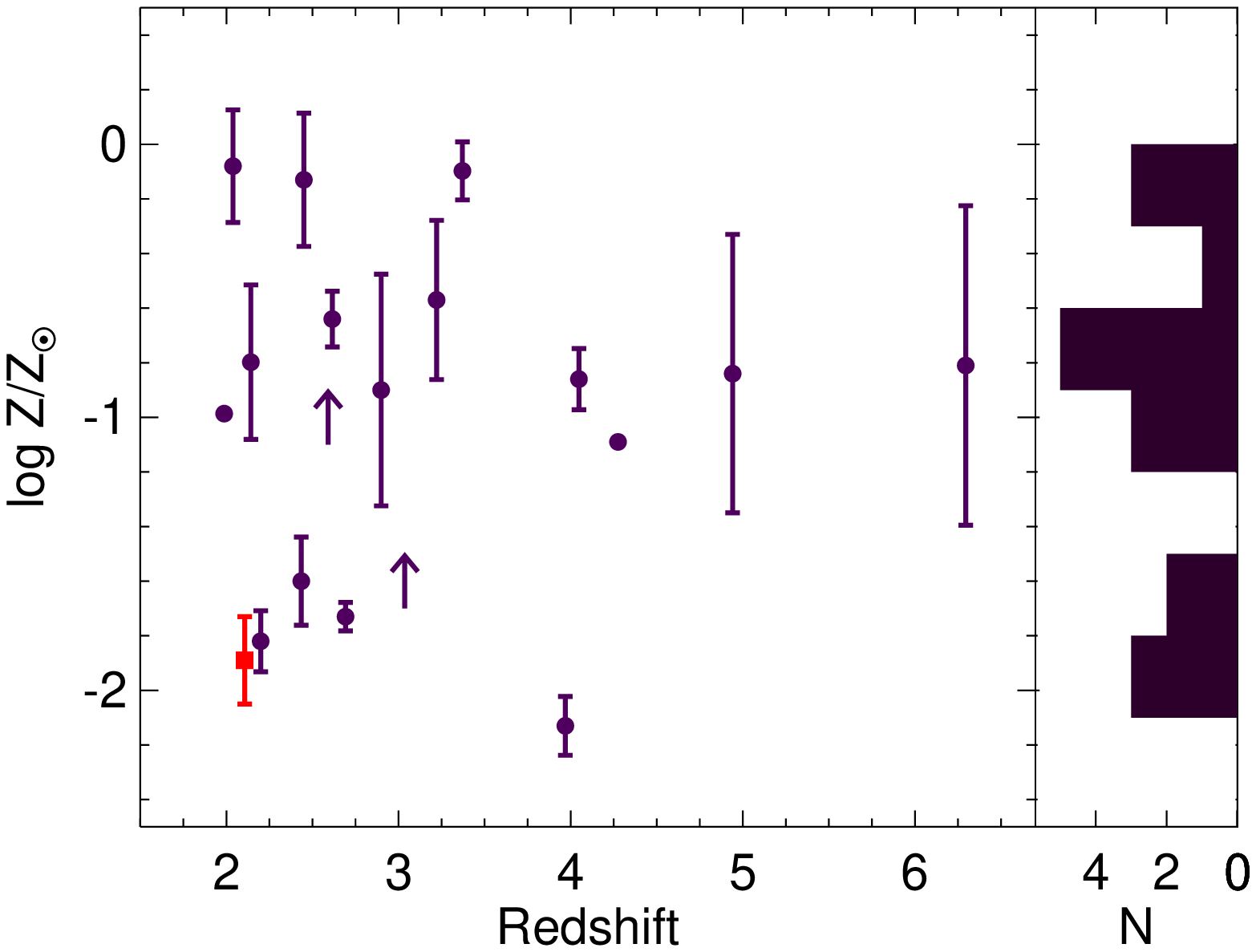}
   \caption{{\it   Left:}   Metallicities    found   in   GRB   Damped
     Lyman-$\alpha$ systems  as function of  redshift.  Filled circles
     mark  metallicities  selected  from  the literature  with  arrows
     indicating lower limits \citep[adopted from ][]{Savaglio:2009aa}.
     The  filled square  shows the  position of  \GRB. The  two points
     without error  bars are for GRBs  030226 and 050505  for which no
     uncertainties for $N_{\rm HI}$  have been reported.  {\it Right:}
     Metallicity         histogram          in         bins         of
     $\Delta\log(Z/Z_{\odot})=0.3$.  Lower limits  have  been excluded
     here.}
   \label{fig:metDistr}
   \end{figure}

With  $\log  (Z/Z_\odot)\approx  -1.9$  \GRB\  has  one  of  the  most
metal-poor DLA  system found in a  GRB so far.  This  is emphasized in
Figure~\ref{fig:metDistr} where we show the metallicity as function of
redshift  for  \GRB\ together  with  18  GRB-DLAs  collected from  the
literature \citep{Prochaska:2007ab,Savaglio:2009aa}.   The addition of
\GRB\ to  the sample strengthens the observation  that the metallicity
spread  in GRB-DLAs at  redshifts 2--3  spans at  least two  orders of
magnitude, from nearly solar \citep[GRB~000926;][]{Savaglio:2003aa} to
1/100th  solar \citep[\GRB  \& GRB  050922C,][]{Prochaska:2008aa}.  At
higher redshifts  the sample becomes  too small to  draw statistically
significant conclusions.  However, no  trend with redshift is visible.
For  the whole sample  ($2.0<z<6.3$), the  mean metallicity  is $<\log
(Z/Z_\odot)> = -1.04\pm0.62$.

%

   The immediate vicinity of a GRB  is shaped by its progenitor and by
   the  energetic explosion  itself.  UV  photons produced  during the
   GRB, or from massive stars  in the star forming region, ionize most
   of   the    circumburst   medium   out   to    a   large   distance
   \citep[e.g.,][]{Perna:1998aa,Waxman:2000aa}.  Empirical constraints
   on  the  ionization  radius  have  been  obtained  from  UV/optical
   absorption line measurements for a  number of bursts and range from
   a  few  tens of  parsec  up to  1.7\,kpc  from  the explosion  site
   \citep[e.g.,][]{Prochaska:2006aa,Dessauges-Zavadsky:2006aa,Vreeswijk:2006aa}.
   While  these values  are significantly  larger than  the ionization
   radii inferred from the  $N_{\rm H}$ equivalent measurements in the
   X-rays      and      the      GRB      ultra-violet      luminosity
   \citep[$\approx3$\,pc;][]{Watson:2007aa},  they  indicate that  the
   low-ionization lines  in the  UV/optical spectra are  not assembled
   from the  immediate vicinity of  the burst, but primarily  from gas
   spread through kpc surrounding regions.

   Metallicity dispersion within a galaxy is well studied in the local
   Universe  \citep[e.g.,][]{Shaver:1983aa,Cioni:2009aa} and similarly
   expected  for  the  host   galaxies  of  GRBs  at  higher  redshift
   \citep[e.g.,][]{Pontzen:2009aa}.   This   would  suggest  that  the
   chemical abundace  derived from a DLA  may differ from  that of the
   molecular  cloud  in which  the  GRB  progenitor  was formed.   The
   extremely  low  metallicity  revealed  by  the DLA  of  \GRB\  does
   therefore not immediately require a similarly metal poor progenitor
   star.  It could have formed in a region with lower metallicity, but
   may also have been more  chemically enriched.  As the same argument
   holds   for  the  the   whole  sample   of  GRB-DLAs   depicted  in
   Figure~\ref{fig:metDistr}, the  observed spread in  metallicity may
   not directly represent a similar spread in the chemical composition
   of the GRB progenitors.  Instead,  the scatter likely arises from a
   combination  of  the  internal  gas-metallicity dispersion  in  the
   individual  host galaxies  as well  as  from the  range of  average
   metallicities of the galaxies included in the host sample.

   Previous  studies have  demonstrated that  GRBs  are preferentially
   formed  in  the  most  luminous  regions of  their  host  galaxies,
   indicating        high        star        formation        activity
   \citep{Fruchter:2006aa,Svensson:2010aa}.   Therefore, the  lines of
   sight probed by their DLAs  are not completely random paths through
   the host  galaxies. Instead,  they trace the  neutral parts  of the
   star forming regions in which  the progenitors were formed, as well
   as molecular clouds, interstellar medium, and halo material.  These
   components will differ in their chemical compositions and different
   sight  lines will  result  in different  net  column densities  and
   abundances.   This   interpretation  is  also   been  supported  by
   hydrodynamical  high-resolution  simulations \citep{Pontzen:2009aa}
   which  found metallicities  of $10^{-2}<Z/Z_\odot<1$  for different
   GRB-DLA sight lines through their host galaxies.

\section{Conclusion}
\GRB\ was  an event of  two extremes.  It  was likely one of  the most
energetic explosions detected so  far and simultaneously showed one of
the most  metal poor  GRB-DLA found until  now.  The  large luminosity
$E_{\gamma}>3.5\times10^{52}$\,erg,  coupled   with  a  bright  slowly
decaying afterglow, allowed a detailed study of the temporal evolution
of the optical transient.  It furthermore enabled high signal-to-noise
spectroscopy  of the  afterglow as  late  as one  day post-burst  with
VLT/FORS2.

With  the spectrum we  confirmed the  redshift of  the host  galaxy of
$z=2.1062\pm0.0004$ and discovered  two intervening absorption systems
at $z=1.946\pm0.001$ and  $z=1.748\pm0.001$. Furthermore, we derived a
neutral    hydrogen    column     density    of    $N_{\rm    HI}    =
10^{21.73\pm0.07}$\,cm$^{-2}$  and a  metallicity of  the  neutral ISM
along the line of sight in the host galaxy of $\log (Z/Z_\odot)\approx
-1.9$.

We close this paper with one reminder. The DLAs found in GRB afterglow
spectra   are  predominantly  probing   the  diverse   conditions  and
sub-structures  within  the   host  galaxies.   They  therefore  allow
important  insight  into  the  gas  and  metallicity  distribution  in
galaxies at  high redshift.  The  evolution of metallicity  and column
densities with  redshift is, however,  challenging to access  with the
current number of  GRB- DLAs.  A much larger sample  of sight lines is
required   to  first   characterize   the  intrinsic   gas-metallicity
dispersions,  and before  GRB-DLAs  can be  used  reliably for  cosmic
chemical evolution studies.


\acknowledgments 

We thank  the referee for the  very helpful comments.  The results are
based  on  observations  made  with  ESO  Telescopes  at  the  Paranal
Observatories under  program 083.D-0903.  We  are grateful to  the ESO
staff, inparticular  Stephane Brillant for the rapid  execution of the
observations.  ARau also thanks Elena Mason for valuable discussion on
the FORS2 data products.  Part of the funding for GROND (both hardware
as well as personnel) was generously granted from the Leibniz-Prize to
Prof.   G.   Hasinger  (DFG  grant  HA~1850/28-1).   SMB  acknowledges
support  of the a  European Union  Marie Curie  European Reintegration
Grant   within   the    7$^{th}$   Program   under   contract   number
PERG04-GA-2008-239176.  TK acknowledges support  by the DFG cluster of
excellence Origin  and Structure  of the Universe.   S.K. acknowledges
support by DFG grant Kl 766/11-3.




\begin{table}
\caption{Log of GROND optical photometry\label{tab:grondLog}}
\begin{tabular}{cccccc}
\hline
\noalign{\smallskip}
$T_{\rm mid} - T_{\rm 0} $ & Exposure &  \multicolumn{4}{c}{Brightness$^a$}  \\  
$[\rm ks]$ & $[\rm s]$  & \multicolumn{4}{c}{mag$_{\rm AB}$$^b$}  \\ 
\hline
 &   & $g^\prime$ & $r^\prime$ & $i^\prime$ & $z^\prime$ \\
\hline
73.16 & 115 & 19.00$\pm$0.03 & 18.70$\pm$0.02 & 18.54$\pm$0.03 & 18.33$\pm$0.03 \\
73.34 & 115 & 18.98$\pm$0.02 & 18.70$\pm$0.02 & 18.53$\pm$0.02 & 18.35$\pm$0.03 \\
73.52 & 115 & 18.98$\pm$0.02 & 18.68$\pm$0.02 & 18.55$\pm$0.02 & 18.35$\pm$0.03 \\
73.44 & 115 & 18.96$\pm$0.02 & 18.68$\pm$0.02 & 18.51$\pm$0.02 & 18.35$\pm$0.03 \\
73.91 & 115 & 18.93$\pm$0.03 & 18.67$\pm$0.02 & 18.53$\pm$0.03 & 18.35$\pm$0.03 \\
74.10 & 115 & 18.95$\pm$0.02 & 18.70$\pm$0.02 & 18.51$\pm$0.02 & 18.36$\pm$0.03 \\
74.30 & 115 & 18.95$\pm$0.02 & 18.70$\pm$0.02 & 18.52$\pm$0.02 & 18.35$\pm$0.03 \\
74.49 & 115 & 18.95$\pm$0.02 & 18.67$\pm$0.02 & 18.50$\pm$0.02 & 18.33$\pm$0.02 \\
74.81 & 375 & 18.94$\pm$0.02 & 18.67$\pm$0.02 & 18.49$\pm$0.02 & 18.31$\pm$0.03 \\
75.26 & 375 & 18.92$\pm$0.02 & 18.67$\pm$0.02 & 18.48$\pm$0.02 & 18.32$\pm$0.02 \\
75.71 & 375 & 18.89$\pm$0.02 & 18.64$\pm$0.02 & 18.46$\pm$0.02 & 18.29$\pm$0.02 \\
76.16 & 375 & 18.90$\pm$0.02 & 18.65$\pm$0.02 & 18.48$\pm$0.02 & 18.31$\pm$0.02 \\
76.77 & 115 & 18.89$\pm$0.02 & 18.65$\pm$0.02 & 18.45$\pm$0.03 & 18.27$\pm$0.03 \\
76.77 & 115 & 18.87$\pm$0.02 & 18.63$\pm$0.02 & 18.47$\pm$0.02 & 18.25$\pm$0.03 \\
76.87 & 115 & 18.88$\pm$0.02 & 18.62$\pm$0.02 & 18.45$\pm$0.02 & 18.27$\pm$0.03 \\
77.06 & 115 & 18.91$\pm$0.02 & 18.62$\pm$0.02 & 18.48$\pm$0.02 & 18.27$\pm$0.03 \\
83.19 & 115 & 18.74$\pm$0.02 & 18.50$\pm$0.02 & 18.36$\pm$0.02 & 18.20$\pm$0.03 \\
83.37 & 115 & 18.75$\pm$0.02 & 18.51$\pm$0.02 & 18.33$\pm$0.02 & 18.21$\pm$0.02 \\
83.56 & 115 & 18.77$\pm$0.02 & 18.52$\pm$0.02 & 18.36$\pm$0.02 & 18.18$\pm$0.02 \\
83.75 & 115 & 18.78$\pm$0.02 & 18.49$\pm$0.02 & 18.36$\pm$0.02 & 18.21$\pm$0.02 \\
83.95 & 115 & 18.74$\pm$0.02 & 18.52$\pm$0.02 & 18.37$\pm$0.02 & 18.20$\pm$0.03 \\
84.14 & 115 & 18.74$\pm$0.02 & 18.52$\pm$0.02 & 18.35$\pm$0.02 & 18.23$\pm$0.02 \\
84.33 & 115 & 18.75$\pm$0.02 & 18.54$\pm$0.02 & 18.37$\pm$0.02 & 18.19$\pm$0.02 \\
84.53 & 115 & 18.74$\pm$0.02 & 18.52$\pm$0.02 & 18.40$\pm$0.02 & 18.18$\pm$0.02 \\
84.73 & 115 & 18.74$\pm$0.02 & 18.53$\pm$0.02 & 18.40$\pm$0.02 & 18.22$\pm$0.03 \\
84.92 & 115 & 18.78$\pm$0.02 & 18.52$\pm$0.02 & 18.38$\pm$0.02 & 18.22$\pm$0.02 \\
85.11 & 115 & 18.76$\pm$0.02 & 18.54$\pm$0.02 & 18.39$\pm$0.02 & 18.22$\pm$0.02 \\
85.31 & 115 & 18.76$\pm$0.02 & 18.52$\pm$0.02 & 18.38$\pm$0.02 & 18.21$\pm$0.02 \\
91.58 & 115 & 18.89$\pm$0.02 & 18.68$\pm$0.02 & 18.49$\pm$0.02 & 18.37$\pm$0.03 \\
91.76 & 115 & 18.92$\pm$0.02 & 18.67$\pm$0.02 & 18.52$\pm$0.02 & 18.37$\pm$0.02 \\
91.95 & 115 & 18.87$\pm$0.02 & 18.69$\pm$0.02 & 18.53$\pm$0.02 & 18.40$\pm$0.03 \\
92.15 & 115 & 18.89$\pm$0.02 & 18.69$\pm$0.02 & 18.51$\pm$0.02 & 18.33$\pm$0.03 \\
92.35 & 115 & 18.89$\pm$0.03 & 18.68$\pm$0.02 & 18.50$\pm$0.03 & 18.36$\pm$0.03 \\
92.54 & 115 & 18.91$\pm$0.02 & 18.69$\pm$0.02 & 18.54$\pm$0.02 & 18.37$\pm$0.03 \\
92.74 & 115 & 18.92$\pm$0.02 & 18.69$\pm$0.02 & 18.54$\pm$0.02 & 18.36$\pm$0.03 \\
92.93 & 115 & 18.94$\pm$0.03 & 18.69$\pm$0.02 & 18.52$\pm$0.03 & 18.36$\pm$0.03 \\
93.14 & 115 & 18.97$\pm$0.03 & 18.74$\pm$0.03 & 18.53$\pm$0.04 & 18.36$\pm$0.04 \\
93.33 & 115 & 18.99$\pm$0.02 & 18.72$\pm$0.03 & 18.55$\pm$0.03 & 18.38$\pm$0.03 \\
93.51 & 115 & 18.90$\pm$0.02 & 18.73$\pm$0.03 & 18.53$\pm$0.03 & 18.38$\pm$0.04 \\
93.71 & 115 & 18.94$\pm$0.03 & 18.72$\pm$0.02 & 18.53$\pm$0.03 & 18.40$\pm$0.03 \\
102.15 & 115 & 19.03$\pm$0.03 & 18.85$\pm$0.03 & 18.63$\pm$0.04 & 18.45$\pm$0.06 \\
102.33 & 115 & 19.08$\pm$0.02 & 18.80$\pm$0.02 & 18.65$\pm$0.03 & 18.49$\pm$0.04 \\
102.52 & 115 & 19.06$\pm$0.02 & 18.83$\pm$0.02 & 18.65$\pm$0.03 & 18.54$\pm$0.05 \\
102.72 & 115 & 19.04$\pm$0.02 & 18.80$\pm$0.02 & 18.65$\pm$0.02 & 18.48$\pm$0.03 \\
102.92 & 115 & 19.05$\pm$0.02 & 18.84$\pm$0.02 & 18.65$\pm$0.03 & 18.47$\pm$0.04 \\
103.11 & 115 & 19.08$\pm$0.02 & 18.81$\pm$0.02 & 18.69$\pm$0.03 & 18.46$\pm$0.04 \\
103.31 & 115 & 19.09$\pm$0.02 & 18.87$\pm$0.02 & 18.68$\pm$0.02 & 18.46$\pm$0.03 \\
103.50 & 115 & 19.07$\pm$0.02 & 18.84$\pm$0.02 & 18.69$\pm$0.02 & 18.47$\pm$0.03 \\
103.71 & 115 & 19.12$\pm$0.02 & 18.86$\pm$0.02 & 18.69$\pm$0.03 & 18.53$\pm$0.04 \\
103.90 & 115 & 19.13$\pm$0.02 & 18.85$\pm$0.02 & 18.68$\pm$0.03 & 18.51$\pm$0.04 \\
104.09 & 115 & 19.09$\pm$0.02 & 18.84$\pm$0.02 & 18.69$\pm$0.02 & 18.53$\pm$0.03 \\
104.29 & 115 & 19.09$\pm$0.02 & 18.83$\pm$0.02 & 18.71$\pm$0.02 & 18.50$\pm$0.03 \\
166.36 & 115 & 19.88$\pm$0.06 & 19.72$\pm$0.04 & 19.50$\pm$0.06 & 19.43$\pm$0.06 \\
166.55 & 115 & 19.90$\pm$0.05 & 19.66$\pm$0.04 & 19.57$\pm$0.05 & 19.39$\pm$0.07 \\
166.74 & 115 & 19.92$\pm$0.05 & 19.69$\pm$0.03 & 19.55$\pm$0.04 & 19.40$\pm$0.06 \\
166.94 & 115 & 19.93$\pm$0.04 & 19.66$\pm$0.03 & 19.53$\pm$0.04 & 19.41$\pm$0.05 \\
167.13 & 115 & 19.88$\pm$0.06 & 19.76$\pm$0.05 & 19.52$\pm$0.06 & 19.36$\pm$0.04 \\
167.32 & 115 & 19.94$\pm$0.07 & 19.70$\pm$0.04 & 19.55$\pm$0.05 & 19.37$\pm$0.07 \\
167.51 & 115 & 19.90$\pm$0.05 & 19.73$\pm$0.03 & 19.60$\pm$0.03 & 19.37$\pm$0.04 \\
167.70 & 115 & 19.89$\pm$0.04 & 19.72$\pm$0.03 & 19.50$\pm$0.05 & 19.40$\pm$0.05 \\
173.77 & 115 & 20.13$\pm$0.05 & 19.85$\pm$0.03 & 19.62$\pm$0.04 & 19.42$\pm$0.06 \\
173.95 & 115 & 20.06$\pm$0.03 & 19.78$\pm$0.02 & 19.68$\pm$0.04 & 19.44$\pm$0.04 \\
174.15 & 115 & 20.08$\pm$0.03 & 19.80$\pm$0.03 & 19.62$\pm$0.03 & 19.41$\pm$0.03 \\
174.35 & 115 & 20.11$\pm$0.04 & 19.80$\pm$0.03 & 19.60$\pm$0.05 & 19.47$\pm$0.05 \\
\hline
\end{tabular}
\end{table}

\begin{table}
\caption{Table~4 continued}
\begin{tabular}{cccccc}
\hline
\noalign{\smallskip}
$T_{\rm mid} - T_{\rm 0} $ & Exposure &  \multicolumn{4}{c}{Brightness$^a$}  \\  
$[\rm ks]$ & $[\rm s]$  & \multicolumn{4}{c}{mag$_{\rm AB}$$^b$}  \\ 
\hline
 &   & $g^\prime$ & $r^\prime$ & $i^\prime$ & $z^\prime$ \\
\hline
174.55 & 115 & 20.09$\pm$0.05 & 19.81$\pm$0.04 & 19.64$\pm$0.04 & 19.44$\pm$0.06 \\
174.73 & 115 & 20.07$\pm$0.03 & 19.82$\pm$0.03 & 19.65$\pm$0.03 & 19.41$\pm$0.05 \\
174.92 & 115 & 20.12$\pm$0.04 & 19.82$\pm$0.03 & 19.63$\pm$0.03 & 19.44$\pm$0.04 \\
175.11 & 115 & 20.14$\pm$0.03 & 19.78$\pm$0.02 & 19.61$\pm$0.03 & 19.47$\pm$0.04 \\
264.66 & 369 & 20.42$\pm$0.05 & 20.15$\pm$0.03 & 20.02$\pm$0.04 & 19.79$\pm$0.05 \\
265.10 & 369 & 20.43$\pm$0.04 & 20.17$\pm$0.03 & 20.04$\pm$0.03 & 19.77$\pm$0.03 \\
265.55 & 369 & 20.45$\pm$0.04 & 20.19$\pm$0.02 & 19.98$\pm$0.03 & 19.79$\pm$0.04 \\
266.00 & 369 & 20.48$\pm$0.04 & 20.21$\pm$0.03 & 19.96$\pm$0.03 & 19.71$\pm$0.03 \\
331.28 & 369 & 20.88$\pm$0.08 & 20.57$\pm$0.05 & 20.35$\pm$0.05 & 20.15$\pm$0.07 \\
331.74 & 369 & 20.85$\pm$0.05 & 20.57$\pm$0.03 & 20.28$\pm$0.03 & 20.09$\pm$0.04 \\
332.19 & 369 & 20.87$\pm$0.05 & 20.55$\pm$0.03 & 20.38$\pm$0.04 & 20.17$\pm$0.04 \\
332.64 & 369 & 20.91$\pm$0.04 & 20.59$\pm$0.03 & 20.34$\pm$0.04 & 20.14$\pm$0.04 \\
355.58 & 369 & 21.05$\pm$0.06 & 20.80$\pm$0.05 & 20.54$\pm$0.05 & 20.34$\pm$0.06 \\
356.02 & 369 & 21.11$\pm$0.05 & 20.83$\pm$0.04 & 20.58$\pm$0.05 & 20.29$\pm$0.05 \\
356.47 & 369 & 21.14$\pm$0.04 & 20.75$\pm$0.03 & 20.56$\pm$0.04 & 20.30$\pm$0.05 \\
356.93 & 369 & 21.09$\pm$0.06 & 20.84$\pm$0.03 & 20.56$\pm$0.04 & 20.40$\pm$0.07 \\
419.04 & 4$\times$369 & 21.15$\pm$0.10 & 20.88$\pm$0.06 & 20.61$\pm$0.07 & 20.45$\pm$0.08 \\
1300.22 & 4$\times$369 & 23.45$\pm$0.12 & 23.16$\pm$0.09 & $>$22.67 & $>$22.47 \\
1635.15 & 4$\times$369 & $>$23.48 & 23.64$\pm$0.12 & $>$22.68 & $>$22.43 \\
1808.91 & 8$\times$369 & 24.13$\pm$0.15 & 23.87$\pm$0.12 & $>$23.17 & $>$22.98 \\
2070.84 & 8$\times$369 & $>$24.24 & 24.02$\pm$0.10 & $>$23.44 & $>$23.32 \\
2840.06 & 12$\times$369 & $>$23.99 & $>$23.98 & $>$23.29 & $>$23.02 \\
\hline
\end{tabular}
\noindent\footnotetext{$^a$Corrected for Galactic foreground reddening}
\footnotetext{$^b$For the SED fitting, the aditional error of the absolute calibration of 0.05 mag was added}
\end{table}


\begin{table}
\caption{Log of GROND near-IR photometry \label{tab:grondJHKphot}}
\begin{center}
\begin{tabular}{ccccc}
\hline
\noalign{\smallskip}
$T_{\rm mid} - T_{\rm 0} $ & Exposure [s] & \multicolumn{3}{c}{Brightness$^a$}  \\  
$[ks]$ & $[s]$ &   \multicolumn{3}{c}{mag$_{\rm AB}$$^b$}  \\ 
\hline
& & $J$ & $H$ & $K_S$ \\
\hline
265.36 & 120 x 10 &  19.44 $\pm$ 0.04 & 19.24 $\pm$ 0.04 & ... \\
331.98 & 120 x 10 &  19.88 $\pm$ 0.04 & 19.52 $\pm$ 0.04 & 19.27 $\pm$ 0.09 \\
356.28 & 120 x 10 &  20.03 $\pm$ 0.04 & 19.77 $\pm$ 0.07 & 19.55 $\pm$ 0.09 \\
419.06 & 120 x 10 &  20.01 $\pm$ 0.06 & 19.80 $\pm$ 0.07 & 19.57 $\pm$ 0.13 \\
\hline
\end{tabular}
\end{center}
\footnotetext{$^a$Corrected for Galactic foreground reddening, but converted to AB magnitudes.}
\footnotetext{$^b$For the SED fitting, the aditional error of the absolute calibration of 0.05 ($J$ and $H$) and 0.07 ($K_S$) mag was added quadratically}
\end{table}

\bibliographystyle{apj}

\end{document}